\documentclass[epj,final]{svjour}

\usepackage{graphicx}
\usepackage{dcolumn}

\newcolumntype{d}{D{.}{.}{3}}
\newcommand{\tbtbt}{$2$$\times$$2$$\times$$2$ }
\newcommand{\obobo}{$1$$\times$$1$$\times$$1$ }
\newcommand{\oh}{\frac{1}{2}}

\begin{document}

\title{Ab Initio Lattice Dynamics and Elastic Constants of ZrC}

\author{P.T. Jochym \and K. Parlinski}
\institute{Institute of Nuclear Physics, 
  ul. Radzikowskiego 152, 31--342 Cracow, Poland
  \mail{jochym@ifj.edu.pl}}

\abstract{ {\it Ab initio} calculations and a direct method are
  applied to derive the phonon dispersion relations and phonon density
  of states for the ZrC crystal. The results are in good agreement
  with neutron scattering data.  The force constants are determined
  from the Hellmann-Feynman forces induced by atomic displacements in
  a \tbtbt supercell.  The elastic constants are found using the
  deformation method and successfully compare with experimental data.
  }


\maketitle


Studies of transition-metal carbides have been carried out several
times. Phenomenological model calculations \cite{weber,gupta,yadav}
managed to reproduce the phonon dispersion relations quite well, but
the interactions were quite sophisticated , some of them including
three-body interactions and double shell models.  Recently, there have
been several studies based on {\em ab initio}
methods\cite{neckel,zhukov,price,haglund}.
The electronic structure, bulk modulus, and elastic constants, as well
as the phonon dispersion relations have been found for TiC, TiN and
TiO compounds by means of first-principles total-energy calculations
\cite{mytic,TiC_GGA}.  Generally, the calculated values show good
agreement with experiments.  This provides a motivation to extend the
investigation to heavier transition metals and to treat zirconium
carbide, ZrC. This crystal is an important material in nuclear energy
technology, since it provides a filling medium for fuel particles. It
is also used for surface hardening and coverage of cathodes of X-ray
sources. In this paper we continue to study the lattice dynamics and
elastic properties of transition-metal compounds. See \cite{mytic} and
references given there for a more detailed description of the method.

The transition-metal carbide compounds, of which ZrC is a
representative, are of considerable scientific and technological
interest because of their striking mechanical properties, extreme
hardness combined with metallic electrical and thermal conductivities.
The ZrC crystal has NaCl structure and it is usually non-stoichiometric,
mainly owing to carbon-vacancy defects.  The phonon dispersion relations
of ZrC have been measured along main symmetry directions by Smith et
al \cite{smithA,smithB}. Elastic constants, in turn, have been
measured for two concentrations of carbon (ZrC$_{0.89}$ and
ZrC$_{0.94}$) by Chang~and Graham \cite{chang}.

In this paper we extend the first-principle calculations to describe
the phonon dispersion curves, phonon density and elastic constants of
ZrC.  The method which is used, is based on the total energy
calculation and {\it Hellmann--Feynman\/} (HF) forces.  The phonon
dispersion relations are calculated by the {\em direct method\/} and
the {\sc Phonon} program
\cite{frank,kresse,parlinski,ackland,sluiter,phonon}, in which the
force constants of the dynamical matrix are derived from HF forces.
Alternatively, one could use the linear--response method
\cite{gonze,giannozzi} for the evaluation of {\em ab initio} phonon
frequencies at a predetermined set of Brillouin zone points.
Elastic constants and bulk modulus are estimated by straightforward
evaluation of energy derivatives with respect to the deformations.

The energies and HF forces of ZrC crystal are calculated by the
method of total energy minimization, using norm-conserving {\em
  pseudopotentials\/} as an approximation for the atomic core--valence
electron interaction \cite{lin,goniakowskiA,payne}.
This method allows to include the pressure in calculations as well as
anharmonic effects.
For the lattice dynamics calculations a \tbtbt supercell with periodic
boundary conditions and 64 atoms was used. For optimizing the
structure and for the direct calculation of stress-strain relations a
\obobo supercell was utilized. The {\it ab initio} total energy
calculations were done with the CASTEP package \cite{castep} and
standard pseudopotentials constructed within LDA approximation and
provided within this package. The Zr pseudopotential treats 4d
electrons as belonging to the valence band.  All pseudopotentials were
parametrized in the reciprocal space with 680 eV cut--off energy.

Tests, which were made with the \obobo supercell and with cut-off
energies of 680~eV and 900~eV, showed that in the first case the
cohesive energy is only 0.01 eV higher, and the lattice constant
changes by only 0.0007~\AA (0.01\%).  Therefore, we used the 680 eV
cut-off energy for all remaining calculations.

The {\em Local Density Approximation\/} (LDA) as well as the {\em
  Generalized Gradient Approximation } (GGA) were used for the
exchange energy term of the valence states of the Hamiltonian
\cite{perdew}. The integration over the Brillouin zone was performed
with weighted summation over wave vectors generated by the
Monkhorst-Pack scheme \cite{pack} using grid spacings from 0.1
\AA$^{-1}$ to 0.04 \AA$^{-1}$ which lead to sets of k-vectors
containing from 4 to 118 wave vectors for the \obobo supercell and
from 1~to 14 wave vectors for the \tbtbt supercell.  The convergence
of the force constants was achieved for grid spacing below
0.05~\AA$^{-1}$.  Hence, we carried on the optimization for a
0.05~\AA$^{-1}$ grid spacing, which leads to 32 and 4 k-points for the
\obobo and the \tbtbt supercells, respectively.  However, we used a 64
wave-vector set for the LDA calculation of the elastic constants.  The
metallic character of the TiC compound, studied in \cite{mytic},
indicated that the use of smearing does not substantially improve the
quality of phonon dispersion relations. Thus, we did not use smearing
in the case of ZrC.
{The use of non--metallic summation of the Brillouin zone is further
  justified by the fact that the electron density of states at the
  Fermi level is small for ZrC and similar compounds: TiC and HfC
  \cite{weber,biltz}, and the related errors are of the same order as
  the errors of the DFT approach itself \cite{mytic}. Namely, they are
  less then 0.05\% for the lattice constant, less then 0.3\% for the
  bulk modulus, and less then 5\% for phonon frequencies.}

\begin{figure}[htbp]
  \leavevmode
  \includegraphics[width=\columnwidth]{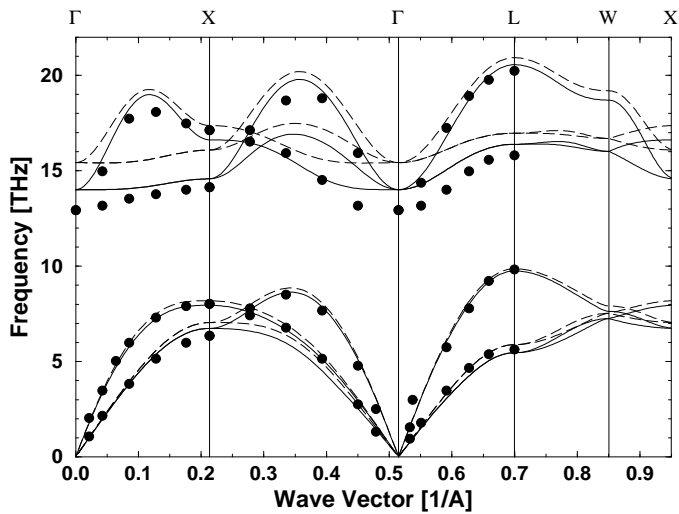}
  \caption{Phonon dispersion relations of ZrC crystal 
    calculated with a~\tbtbt supercell within GGA (full line) and LDA
    (dashed line) approximations. The experimental points are taken
    from Ref.~\cite{smithA,smithB}.}
    \label{fig:big_cell}
\end{figure}

The minimization of the total energy leads to the equilibrium lattice
constant of the stoichiometric ZrC $a=4.691$\AA{} for LDA and
$a=4.695$\AA{} for GGA. These values could be compared with the
experimental values of $a=4.6994$\AA{} for ZrC$_{0.94}$ and
$a=4.7004$\AA{} for ZrC$_{0.89}$\cite{chang}.  
The phonon dispersion relations, which correspond to temperature T=0,
are shown in Fig.~\ref{fig:big_cell}. These are compared with
experimental phonon frequencies measured by inelastic neutron
scattering \cite{smithA,smithB}.
For GGA, the calculated force constants for all 10 coordination
spheres of the \tbtbt supercell are presented in
Table~\ref{tab:fc}.

\begin{table*}[htbp]
  \begin{center}
    \begin{tabular}{lccdddddd}
      \hline\hline
      \multicolumn{1}{c}{Distance (\AA)} & From & To              
      & xx     & yy     & zz     & yz     & xz     & xy     \\
      \hline
      \multicolumn{9}{c}{} \\
      0.0    & C$_{(0,0,0)}$        & C$_{(0,0,0)}$        & 219.36 & 219.36 & 219.36 &        &        &        \\
      0.0    & Zr$_{(\oh,\oh,\oh)}$ & Zr$_{(\oh,\oh,\oh)}$ & 284.67 & 284.67 & 284.67 &        &        &        \\
      2.3447 & C$_{(0,0,0)}$        & Zr$_{(\oh,0,0)}$     & -24.20 & -23.82 & -23.82 &        &        &        \\
      3.3159 & C$_{(0,0,0)}$        & C$_{(\oh,\oh,0)}$    &  -5.77 &  -5.77 &   0.64 &        &        & -10.41 \\
      3.3159 & Zr$_{(\oh,\oh,\oh)}$ & Zr$_{(1,1,\oh)}$     & -15.67 & -15.67 &  -3.68 &        &        & -31.66 \\
      4.0611 & C$_{(0,0,0)}$        & Zr$_{(\oh,\oh,\oh)}$ &  -1.90 &  -1.90 &  -1.90 &  -1.20 &  -1.20 &  -1.20 \\
      4.6894 & C$_{(0,0,0)}$        & C$_{(1,0,0)}$        & -21.97 &   1.02 &   1.02 &        &        &        \\
      4.6894 & Zr$_{(\oh,\oh,\oh)}$ & Zr$_{(1\oh,\oh,\oh)}$& -17.47 &   3.82 &   3.82 &        &        &        \\
      5.2429 & C$_{(0,0,0)}$        & Zr$_{(1,\oh,0)}$     &   2.63 &  -0.01 &   0.53 &        &        &        \\
      5.7433 & C$_{(0,0,0)}$        & C$_{(1,\oh,\oh)}$    &   0.37 &   0.36 &   0.36 &  -0.15 &  -0.02 &  -0.02 \\
      5.7433 & Zr$_{(\oh,\oh,\oh)}$ & Zr$_{(1\oh,1,1)}$    &   0.39 &   0.28 &   0.28 &   0.41 &        &        \\
      6.6318 & C$_{(0,0,0)}$ &  C$_{(1,1,0)}$              &  -0.94 &  -0.94 &  -0.05 &        &        &        \\
      6.6318 & Zr$_{(\oh,\oh,\oh)}$ & Zr$_{(1\oh,1\oh,\oh)}$&  0.52 &   0.52 &  -0.06 &        &        &        \\
      7.0341 & C$_{(0,0,0)}$ &  Zr$_{(1,1,\oh)}$           &  -0.13 &  -0.13 &  -0.09 &        &        &        \\
      8.1223 & C$_{(0,0,0)}$ &  C$_{(1,1,1)}$              &  -0.03 &  -0.03 &  -0.03 &        &        &        \\
      8.1223 & Zr$_{(\oh,\oh,\oh)}$ & Zr$_{(1\oh,1\oh,1\oh)}$&-0.01 &  -0.01 &  -0.01 &        &        &        \\
      \multicolumn{9}{c}{} \\
      \hline\hline
    \end{tabular}
  \end{center}
  \caption{Values of the non-zero elements of the force-constant matrices 
    derived from a~\tbtbt supercell in GGA (in N/m). 
    All force-constant matrices are symmetric.}
      \label{tab:fc}
\end{table*}

From Fig.~\ref{fig:big_cell} one sees that the experimental data at
the $\Gamma$ point are about $\Delta\omega = 0.05 \omega$ lower than
the calculated values. A reason for this could be that the
experimental values are measured at room temperature and that
anharmonic effects diminish the phonon frequencies.  Another reason
may relate to the non-stoichiometric concentration of carbon, which
diminishes the average phonon frequencies as well. In other {\em
  ab-initio} calculations for alkali metals\cite{frank}, graphite and
diamond\cite{kresse} the experimental phonon points also are at
frequencies lower than calculated ones. On the same
Fig.~\ref{fig:big_cell} we see the influence of the different
exchange-correlation energy approximations on phonon frequencies. In
this case, the LDA appears to be far less accurate than the GGA.

\begin{figure}[tb]
  \leavevmode 
  \includegraphics[width=\columnwidth]{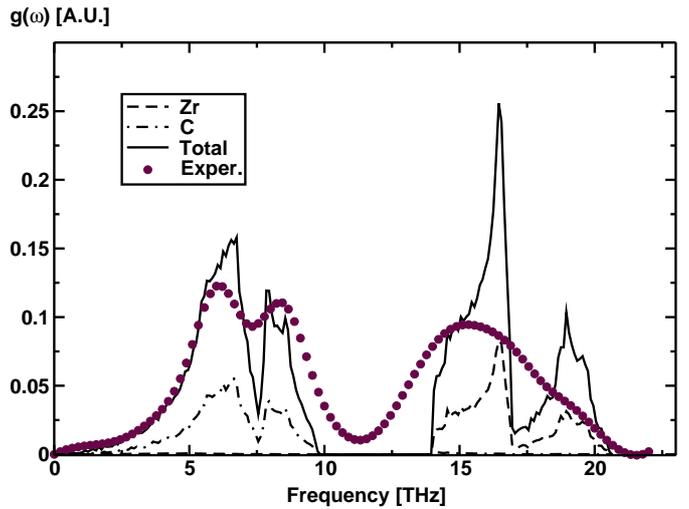} 
 \caption{Phonon density of states $g(\omega)$ of the ZrC crystal and 
   partial phonon density of states $g_{i,\alpha }(\omega)$, where
   $\alpha=$ C or Zr and $i=x,$ $y$ or $z$ are the displacement
   directions. These are the GGA results.  Experimental data taken from
   Ref.~\cite{gompf}.  }
  \label{fig:statdens} 
\end{figure}

By sampling the dynamical matrix at many wave vectors, one can
calculate the phonon density of states $g(\omega )$, and the partial
phonon density of states $g_{x, Zr}(\omega )$ and $g_{x, C}(\omega )$.
The $g(\omega )$ describes the number of phonon frequencies in an
interval around $\omega $, while $g_{x, Zr}$ and $g_{x, C}$ specify
the number of phonon frequencies in an interval around $\omega $, but
only taking into account vibrations of Zr and C atoms, respectively.
The density-of-state functions calculated from the GGA data are shown
in Fig.~\ref{fig:statdens}.  They are conventionally normalized to
$\int g(\omega ) d\omega = 1$ and $\int g_{x,\alpha }(\omega ) d\omega
= 1/6$, where $\alpha =$ Zr or C.  The curves show that motions within
acoustic dispersion curves are almost entirely due to Zr atoms.  The
sum of the zirconium density of states ($g_{x, Zr}(\omega ) + g_{y,
  Zr}(\omega ) + g_{z, Zr}(\omega )$) fits the total density of states
$g(\omega )$ below 10 THz. The part of $g(\omega )$ above the gap is
mainly due to vibrations of C atoms.  Thus, ZrC forms quite a special
crystal, in which the heavy Zr atoms, form a frame for the elastic
motion, and the C atoms vibrate within the optical modes.  We see in
Table~\ref{tab:fc}, that the magnitudes of the force constants between
Zr--C and between Zr--Zr and C--C at a similar distance remain of the
same order.  Thus, ZrC does not consist of two weakly bounded
subsystems.  The situation is similar in TiC \cite{mytic}. In
Fig.~\ref{fig:statdens} the experimentally determined phonon density
of states taken from Ref.\cite{gompf} is also shown.  The agreement
within the acoustic region is quite good. In the optical region the
experimental resolution of $\approx 2.5$~THz and the non-stoichiometry
of carbon lead to considerable broadening.

\begin{table}[htbp]
  \begin{tabular}{lD{.}{.}{4}ddd}
    \hline\hline
    Result & 
    \multicolumn{1}{c}{$B$} & 
    \multicolumn{1}{c}{$c_{11}$} & 
    \multicolumn{1}{c}{$c_{12}$} & 
    \multicolumn{1}{c}{$c_{44}$} \\
    \hline
    Expt ZrC$_{0.94}$ \cite{smithA}
              & (2.231) & 4.720 & 0.987 & 1.593 \\
    Expt ZrC$_{0.89}$ \cite{smithA}
              & (2.225) & 4.682 & 0.997 & 1.573 \\
    ZrC extrapolated & (2.238) & 4.766 & 0.975 & 1.617 \\
    Calc. LDA & (2.270) & 4.716 & 1.047 & 1.348 \\
    Calc. LDA (energ.)&  2.295  &       &       &       \\
    Calc. GGA & (2.279) & 4.802 & 1.018 & 1.697 \\
    Calc. GGA (energ.)&  2.32  &       &       &       \\
    \hline\hline
  \end{tabular}
  \caption{Experimental and calculated bulk modulus $B$ 
    and elastic constants $c_{ij}$
    of the ZrC crystal, in units of 
    $10^{11}{\rm Nm}^{-2} = 1 {\rm MBar}$.
    Values in parentheses are calculated from 
    $B=\frac{1}{3}(c_{11}+2c_{12})$.}
    \label{tab:elastic}
\end{table}

The bulk modulus was calculated in the \obobo supercell from the
relation of pressure to the lattice constant. These data were fitted
to a fourth order polynomial.  Hence, the derivative ${\partial
  P}/{\partial V}$, and the bulk modulus $B$ were calculated from the
relation: $B=-V({\partial P}/{\partial V})$, where $V$ is the volume
of the ZrC crystallographic unit cell.
For LDA and GGA, the $B$ values are given in Table~\ref{tab:elastic}
and they agree quite well with experimental data. We also derived the
bulk modulus from the energy-deformation relationship and the
stress--strain relations, and the last proved to be more accurate than
differentiation of the total energy curve (see
Table~\ref{tab:elastic}).

We used stress--strain relations to obtain the elastic constants. The
calculations of $c_{11}$, $c_{12}$, $c_{44}$ elastic constants
involved elongation and shear deformations.  Deformations from $0.5\%$
to $3\%$ in length and from $1$ to $5$ degrees in angle are used.  The
results of small and large deformations are consistent.  The deformed
lattices have space groups Fm$\bar3$m, I4/mmm and I/mmm, for bulk
expansion, elongation along z, and shear modes, respectively.  In all
these deformed lattices all atoms remain in high-symmetry sites so
that the relaxation of the internal degrees of freedom during the
supercell deformations are not necessary. The calculated elastic
constants are compared with the experimental data in
Table~\ref{tab:elastic}. Generally they are in good agreement with
experiments. The $c_{12}$, $c_{44}$ elastic constants are better
reproduced by the GGA, but also $c_{11}$ fits better with GGA, when
experimental data are extrapolated to the stoichiometric ZrC (third
row in Table~\ref{tab:elastic}).

\medskip

In summary, we have shown that the {\em ab initio} calculations of HF
forces together with the direct method lead to a~satisfactory
description of phonons and elastic constants in the ZrC crystal.  The
phonon dispersion relations and the density of states fit well to
neutron experimental data. The GGA offered better phonon frequencies
and elastic constants.

\begin{acknowledgement}
  The use of facilities of the ACC ``Cyfronet'', Cracow, where the
  calculations have been done, are kindly acknowledged.  This work was
  partially supported by the Polish State Committee of Scientific
  Research (KBN), grant No 2~PO3B~004~14, and computational grant No.
  KBN/SGI\_ORIGIN\_2000/IFJ/128/1998.
\end{acknowledgement}



\begin{thebibliography}{99}
\bibitem{castep} CASTEP is a part of $Cerius^2$ system developed
        by Molecular Simulations.
\end{thebibliography}
\end{document}